\global\def\draftcontrol{0}

   \def\versionno{ cqcd }

\catcode`\@=11

\expandafter\ifx\csname draftcontrol\endcsname\relax\global\def\draftcontrol{0}
\fi

{\count255=\time\divide\count255 by 60
\xdef\hourmin{\number\count255}
\multiply\count255 by-60\advance\count255 by\time
\xdef\hourmin{\hourmin:\ifnum\count255<10 0\fi\the\count255}}
\def\draftdate{\number\month/\number\day/\number\year\ \ \ \hourmin }

\newcommand\makepapertitle{\par
  \begingroup
    \renewcommand\thefootnote{\@fnsymbol\c@footnote}%
    \def\@makefnmark{\rlap{\@textsuperscript{\normalfont\@thefnmark}}}%
    \long\def\@makefntext##1{\parindent 1em\noindent
            \hb@xt@1.8em{%
                \hss\@textsuperscript{\normalfont\@thefnmark}}##1}%
     \newpage
     \global\@topnum\z@   
     \@makepapertitle
     \thispagestyle{empty}\@thanks
  \endgroup
  \setcounter{footnote}{0}%
  \global\let\thanks\relax
  \global\let\makepapertitle\relax
  \global\let\@makepapertitle\relax
  \global\let\@thanks\@empty
  \global\let\@author\@empty
  \global\let\@date\@empty
  \global\let\@title\@empty
  \global\let\title\relax
  \global\let\author\relax
  \global\let\date\relax
  \global\let\and\relax
  \def\version{\let\version\@version\@gobble}
}
\def\@makepapertitle{%
  \newpage
   \ifnum\draftcontrol=1 {}
   \version\versionno
   \vskip 3em%
   \else
   \hfill\hbox to 3cm {\parbox{4cm}{\@pubnum}\hss}%
   \vskip 3em%
   \fi
   \begin{center}%
   \let \footnote \thanks
     {\LARGE {\@title}}%
     \vskip 1.5em%
     {\normalsize
       \lineskip .5em%
       \begin{tabular}[t]{c}%
         \@author
       \end{tabular}\par}%
     \vskip 1.5em%
     {\@bstract}%
     \end{center}%
     \vskip 1.5em
     \@date%
   \par
}

\gdef\@pubnum{}
\def\pubnum#1{%
  \gdef\@pubnum{#1}}

\gdef\@bstract{}
\def\Abstract#1{%
  \gdef\@bstract{%
   \parbox{\textwidth-0pc}{%
   \centerline{\bf Abstract}\penalty1000%
\kern.2cm%
\noindent
\renewcommand\baselinestretch{1.0}%
{#1}}}
}

\def\ps@paper{\let\@mkboth\@gobbletwo%
     \ifnum\draftcontrol=1
    \def\@oddfoot{\hbox to \textwidth{\tiny \versionno \hfil\tiny\draftdate}%
    \hskip -\textwidth \hbox to \textwidth{\hfil\rm\thepage\hfil}}%
     \else\def\@oddfoot{\hbox to \textwidth{\hfil\rm\thepage\hfil}}
     \fi
     \let\@evenfoot\@oddfoot
}

\def\body{\clearpage
          \pagestyle{paper}
    }

\def\@version#1{\ifnum\draftcontrol=1
\typeout{}\typeout{#1}\typeout{}
\vskip3mm\centerline{\hbox{\fbox{\normalsize{\tt DRAFT -- #1 -- }
                   {\draftdate}}}}\vskip3mm
\fi}
\let\version\@version
\long\def\eqlabel#1{\ifnum\draftcontrol=1
                    \tag@false  
                    \tag*{(\theequation) \hbox to -0.2cm{\hspace{0cm}\small{#1}\hss}}
                    \refstepcounter{equation}
                    \edef\@currentlabel{\theequation}
                    \ltx@label{#1}          
                    \else
                    \label{#1}
                    \fi
                    }
\let\st@bibitem\@bibitem
\let\st@lbibitem\@lbibitem
\ifnum\draftcontrol=1
  \def\@bibitem#1{%
    \st@bibitem{#1}\a@@label{#1}\ignorespaces}
  \def\@lbibitem[#1]#2{%
    \st@lbibitem[#1]{#2}\a@@label{#2}\ignorespaces}
  \def\a@@label#1{%
    \gdef\a@lab{\smash{\normalfont\small#1}}
    \ifvmode
      \if@inlabel
        \global\setbox\@labels\hbox{%
          \llap{\a@lab\let\a@lab\relax
                \kern\@totalleftmargin\kern\marginparsep}%
          \box\@labels}%
      \fi
    \fi}
\fi

\documentclass[12pt,letterpaper]{article}

\usepackage{amsmath,amssymb,array,calc,epsfig,rotating,bm}
\usepackage[sort]{cite}

\ifnum\draftcontrol=1
\fi

\renewcommand\baselinestretch{1.25}
\renewcommand\section{\@startsection {section}{1}{\z@}%
                                   {-3.5ex \@plus -1ex \@minus -.2ex}%
                                   {2.3ex \@plus.2ex}%
                                   {\normalfont\large\bfseries}}
\renewcommand\subsection{\@startsection{subsection}{2}{\z@}%
                                   {-3.25ex\@plus -1ex \@minus -.2ex}%
                                   {1.5ex \@plus .2ex}%
                                   {\normalfont\normalsize\bfseries}}
\renewcommand\subsubsection{\@startsection{subsubsection}{3}{\z@}%
                                   {-3.25ex\@plus -1ex \@minus -.2ex}%
                                   {1.5ex \@plus .2ex}%
                                   {\normalfont\normalsize\it}}
\renewcommand\paragraph{\@startsection{paragraph}{4}{\z@}%
                                   {-3.25ex\@plus -1ex \@minus -.2ex}%
                                   {1.5ex \@plus .2ex}%
                                   {\normalfont\normalsize\bf}}

\numberwithin{equation}{section}

\def\revise#1       {\raisebox{-0em}{\rule{3pt}{1em}}%
                     \marginpar{\raisebox{.5em}{\vrule width3pt\
                     \vrule width0pt height 0pt depth0.5em
                     \hbox to 0cm{\hspace{0cm}{%
                     \parbox[t]{4em}{\raggedright\footnotesize{#1}}}\hss}}}}

\def\cale         {{\cal E}}

\def\call         {{\cal L}}

\def\caln         {{\cal N}}
\def\calo         {{\cal O}}

\def\sqr#1#2{{\vcenter{\vbox{\hrule height.#2pt
 \hbox{\vrule width.#2pt height#1pt \kern#1pt
 \vrule width.#2pt}\hrule height.#2pt}}}}

\def\a{\alpha}
\def\b{\beta}

\newcommand{\beq}{\begin{equation}}
\newcommand{\eeq}{\end{equation}}
\newcommand{\beqa}{\begin{eqnarray}}
\newcommand{\eeqa}{\end{eqnarray}}
\newcommand{\beqar}{\begin{eqnarray*}}
\newcommand{\eeqar}{\end{eqnarray*}}

\newcommand{\reef}[1]{(\ref{#1})}
\renewcommand{\eqref}[1]{(\ref{#1})}

\newcommand{\eg}{{\it e.g.,}\ }
\newcommand{\ie}{{\it i.e.,}\ }

\newcommand{\mt}[1]{\textrm{\tiny #1}}

\newcommand{\lp}{\ell_{\mt P}}
\newcommand{\veps}{\varepsilon}

\def\a{\alpha}

\def\t{\delta}

\def\dd{{\delta}}

\catcode`\@=12

\begin{document}


\title{\bf sQGP as hCFT}
\pubnum{UWO-TH-09/14}


\author{
Alex Buchel,$ ^{1,2}$  Michal P. Heller$ ^3$ and Robert C. Myers$ ^{1}$ \\[0.4cm]
\it $ ^1$Perimeter Institute for Theoretical Physics\\
\it Waterloo, Ontario N2L 2Y5, Canada\\[.5em]
 \it $ ^2$Department of Applied Mathematics\\
 \it University of Western Ontario\\
\it London, Ontario N6A 5B7, Canada\\[.5em]
\it $ ^3$ Institute of Physics, Jagellonian University\\
\it Reymonta 4, 30-059 Krakow, Poland
 }

\Abstract{We examine the proposal to make quantitative comparisons
between the strongly coupled quark-gluon plasma and holographic
descriptions of conformal field theory. In this note, we calculate
corrections to certain transport coefficients appearing in
second-order hydrodynamics from higher curvature terms to the dual
gravity theory. We also clarify how these results might be
consistently applied in comparisons with the sQGP.}

\makepapertitle

\body

\version\versionno
\tableofcontents

\section{Introduction}

Recent experimental results from the Relativistic Heavy Ion Collider
(RHIC) have revealed a new phase of nuclear matter, known as the
strongly coupled quark-gluon plasma (sQGP) \cite{shuryak}. At the
same time, the AdS/CFT correspondence has matured into a powerful
tool to study thermal and hydrodynamic properties of strongly
coupled gauge theories \cite{review1}. Even though QCD is not (yet)
a gauge theory which has a controllable description in the framework
of a rigorous gauge theory/string theory correspondence, some of its
properties just above the deconfinement phase transition are
remarkably similar to those found for plasmas in holographic
conformal field theory (hCFT) \cite{sym,qm09}. This hints that
certain aspects of the physics may be universal and so may be
accessible in a general AdS/CFT framework. The canonical model is
most notably the $\caln=4$ $SU(N)$ supersymmetric Yang-Mills (SYM)
theory in the planar limit and for large 't Hooft coupling
\cite{juan}. The holographic dual for this and a large class of
superconformal gauge theories is simply Einstein gravity in AdS$_5$
\cite{bmps}. By adding higher curvature interactions to the dual
gravity theory, one expands the range of physical characteristics of
the gauge theory and in \cite{bms} it was suggested that this may
provide a phenomenological approach for quantitative comparisons
between hCFT plasmas and the sQGP. An interesting step in this
direction was made in \cite{comp}.

In this note, we further examine the proposal for such quantitative
comparisons. First in section \ref{hcft}, we briefly describe the
framework of the dual gravity theory with higher curvature
corrections and also present various results for thermal and
hydrodynamic properties of the corresponding conformal plasma. While
much of these results are collected from the previous literature,
the $R^2$ and $R^3$ corrections to the relaxation time $\tau_\Pi$
and second-order transport coefficient $\lambda_1$ are new. As the
calculations are straightforward, we only present results here. In
section \ref{disc}, we translate the results from a description in
terms of the couplings in the dual gravity theory to one in terms of
physical parameters which directly characterize the underlying CFT.
Further, we discuss how these results can be consistently applied in
comparing the results of the holographic calculations to data for
the sQGP. In particular, we distinguish scenarios with and without
exactly marginal couplings.

\section{Holographic conformal hydrodynamics} \label{hcft}

Following \cite{bms}, we consider the holographic description of a
strongly coupled CFT with the following higher curvature
corrections:
\begin{equation}
I=\frac{1}{2\lp^3}\int d^5x \sqrt{-g}\biggl[\frac{12}{L^2} + R +
L^2\,\alpha_1\, C_{abcd}C^{abcd} +L^4\,\a_2\,
C_{ab}{}^{cd}C_{cd}{}^{ef}C_{ef}{}^{ab}  +L^6\, \a_3\, W(C)
\biggr]\,, \eqlabel{act2}
\end{equation}
where $C_{abcd}$ is a five-dimensional Weyl tensor. $W(C)\sim C^4$
is a particular quartic contraction of $C_{abcd}$ which naturally
arises in type IIB supergravity \cite{bmps}. We emphasize that we
assume that $\alpha_n\ll1$ which allows us to treat the higher
curvature couplings perturbatively. In the following, we work to
linear order in $\alpha_{2,3}$ while keeping corrections to second
order in $\alpha_1$. We discuss the rationale for this approach in
section \ref{disc}.

In general, in constructing the gravitational action \reef{act2},
one might have added many more higher curvature interactions,
however, within the present perturbative framework, most of these
can be removed by field redefinitions without affecting the final
physical results, as explained in detail in \cite{bms}. Hence our
perturbative results are completely general for a gravitational
action with interactions quadratic and cubic in
curvatures.\footnote{While the analysis of \cite{bd} introduces two
$R^3$ interactions, a certain combination of these terms vanishes by
a Schouten identity after allowing for field redefinitions.} In
fact, in complete generality, there are five independent
interactions that would appear at order $C^4$ \cite{tba2}. Hence our
analysis is specialized at this order by focusing on the particular
supergravity term in \reef{act2}.

A regular black brane solution  describes the equilibrium
thermodynamics of the CFT plasma. It is straightforward to
incorporate the higher curvature corrections to the Einstein
equations \cite{gkt,kp,bd} and to order in
$\calo(\a_1^2,\a_2,\a_3)$, the equilibrium pressure is given by
\begin{equation}
P=\frac{\pi^4}{2}  \frac{L^3}{\lp^3} \,T^4
\biggl(1+18\a_1+24\a_1^2+24\a_2+15\a_3\biggr) \eqlabel{p1}\,.
\end{equation}
This expression can also be related to the energy density $\veps$ or
entropy density $s$, using standard relations that apply for any
CFT, \ie $\veps=3P$ and $\veps=\frac{3}{4}Ts$ (in the absence of a
chemical potential).

The shear viscosity $\eta$ and the relaxation time $\tau_\Pi$ of the
CFT plasma can be extracted from the two-point boundary
stress-energy correlation functions in the black brane background
\cite{review1,higher}, while the second-order coefficient
$\lambda_1$ can be extracted by studying the holographic dual of the
boost-invariant expansion of plasma \cite{janik}. The leading order
results for $\{\eta,\tau_{\Pi},\lambda_1\}$ were obtained in
\cite{l1,j3,higher} and the $\calo(\a_3)$ corrections were studied
in \cite{f1,check,bp1}. The $\calo(\a_1)$ correction to the shear
viscosity were first considered in \cite{kp} and these results were
extended to include $\calo(\a_1^2,\a_2)$ corrections in \cite{bd}.
It is straightforward to repeat the computations of \cite{bp1} with
the effective action \eqref{act2} to determine corresponding
$\calo(\a_1^2,\a_2)$ corrections to the second-order transport
coefficients $\tau_\Pi$ and $\lambda_1$.\footnote{In the context of
Gauss-Bonnet gravity, numerical calculations of $\a_1$-modifications
to $\tau_\Pi$ were made in \cite{relax}.} We do not include any
details of the analysis here but only present the final results.
Hence, to order $\calo(\a_1^2,\a_2,\a_3)$,
\begin{equation}
\begin{split}
&\frac \eta s=\frac{1}{4\pi}\biggl(1-8\ \a_1+112\ \a_1^2-384\
\a_2+120\ \a_3\biggr)\,,\\
&\tau_\Pi T=\frac{1}{2\pi} \biggl(2-\ln2-11\ \a_1-125\ \a_1^2-104\
\a_2 +\frac{375}{2}\ \a_3\biggr)\,,\\
&\frac {{\lambda}_1 T}{\eta}= \frac{1}{2\pi}\biggl(1-2\ \a_1-146\
\a_1^2-32\ \a_2+215\ \a_3\biggr)\,.
\end{split}
\eqlabel{transport}
\end{equation}

As discussed in the following section, the above results can also be
expressed in terms of physical parameters of the dual CFT. Towards
this end, we compute the two central charges in the CFT dual to
\reef{act2} using the the holographic trace anomaly \cite{renorm1}:
\begin{equation}
{a}=\pi^2\frac{L^3}{\lp^3}\,,\qquad
{c}=\pi^2\frac{L^3}{\lp^3}\bigl(1+8\a_1 \bigr)\,. \eqlabel{ca}
\end{equation}
The $R^2$ and $R^3$ contributions to the central charges were
considered in \cite{two} and \cite{bd}, respectively. However, in
contrast with, \eg \reef{transport} where we have the leading terms
in an (infinite) expansion, we emphasize that these results
\reef{ca} have no higher order corrections with the effective action
\reef{act2}. The fact that \reef{ca} is exact occurs because we have
parameterized the higher curvature corrections only in terms of the
Weyl tensor, which vanishes in the AdS$_5$ background.\footnote{We
thank Aninda Sinha for discussions on this point.}

\section{Discussion}\label{disc}

Working perturbatively in the gravitational couplings in the
effective action \reef{act2}, we have the results for a number of
interesting properties of strongly coupled plasmas in the dual
conformal field theory. Of course, these expressions for the
pressure \reef{p1} and the transport coefficients \reef{transport}
are given in terms of the gravitational couplings $\alpha_n$, as
well as the dimensionless ratio $L/\lp$. As such, these results must
also be specified as arising from our particular presentation of the
effective action \reef{act2}. In general, field redefinitions allow
us to modify the form of the effective action but they will also
change the precise form of these expressions, at the order that we
have presented the results in the previous section. However, as we
now discuss, this ambiguity can be avoided by parameterizing the
results in terms of physical parameters of the underlying CFT.

As alluded to above, two useful parameters which characterize any
four-dimensional CFT are the central charges, $a$ and $c$. Hence
given the results of the holographic trace anomaly \reef{ca}, it is
convenient to replace:
\begin{equation}
\frac{L^3}{\lp^3}=\frac{a}{\pi^2}\,,\qquad
\a_1=\frac{1}{8}\,\frac{c-a}{a}\equiv\frac \delta 8\,. \eqlabel{ca2}
\end{equation}
Again, we emphasize that these expressions are exact and do not
receive further perturbative corrections with our effective action
\reef{act2}.\footnote{Note that our present definition of $\delta$
differs slightly from that in \cite{bms}. There we had
$\delta'=(c-a)/c\simeq 8\a_1-64\a_1^2+\calo(\a_1^3)$. We return to
this choice of parameters below.}

As the corresponding interaction is cubic in the Weyl tensor, $\a_2$
naturally plays a role in defining the three-point function of the
stress tensor in the dual CFT. In general, this three-point function
depends on three independent constants \cite{osborn}. In fact, the
central charges, $a$ and $c$, each corresponds to a certain linear
combination of these parameters. Recently, it was also shown in
\cite{hm} that these constants in the three-point function also
define two new parameters with a clear physical significance in the
CFT. They considered an ``experiment'' in which the energy flux was
measured at null infinity after a local disturbance was created the
insertion of the stress tensor $\calo=T^{ij}\epsilon_{ij}$. The
energy flux escaping at null infinity in the direction indicated by
the unit vector $\vec{n}$ is then \cite{hm}
\begin{equation}
\langle \cale(\vec{n})\rangle_{\calo} =\frac{E}{4\pi}\biggl[1+t_2
\left(\frac{\epsilon_{ij}^*\epsilon_{i\ell}n_in_j}
{\epsilon^*_{ij}\epsilon_{ij}}-\frac 13\right)+t_4
\left(\frac{|\epsilon_{ij}n_in_j|^2}{\epsilon^*_{ij}\epsilon_{ij}}-\frac{2}{15}
\right)\biggr]\,,
 \eqlabel{t2t4def}
\end{equation}
where $E$ is the total energy of the state. The two constants, $t_2$
and $t_4$, can be used to characterize the underlying CFT.

However, recall that the three-point function contains only three
independent parameters which go into defining the four constants:
$a$, $c$, $t_2$ and $t_4$. Hence the latter are not all independent
and rather satisfy the relation \cite{tba}
\begin{equation}
\frac{a}{c}=1-\frac{1}{6}t_2+\frac{4}{45}t_4\,. \eqlabel{t2t4}
\end{equation}
Hence we keep only $t_4$ to characterize the CFT as it is most
naturally connected to the cubic curvature interaction in the dual
gravity action \reef{act2}. To leading order, one finds
\begin{equation}
t_4=4320\,\a_2+\calo(\a_1\a_2,\a_3^2)\,. \eqlabel{t4def}
\end{equation}
We should comment that this result was originally calculated in the
absence of any quartic curvature interactions \cite{hm}, however,
there is no contribution linear in $\a_3$ by essentially the same
reasoning presented in discussing \reef{ca}. The key point is that
our higher curvature corrections in \reef{act2} are written in terms
of the Weyl tensor, which vanishes in the AdS$_5$ background. Hence
the quartic term $W(C)$ cannot contribute the three-point function
and $t_4$ at linear order. Further $t_4$ does not receive a
contribution at order $\a_1^2$, as can be observed by noting that
$t_4$ vanishes identically when the gravitational dual contains only
curvature-squared corrections \cite{tba,diego}. Finally, we add that
$t_4$ has the interesting property that it vanishes when the
underlying CFT is supersymmetric \cite{hm}.

Turning to $\a_3$, one would have to find an analogous parameter
that characterizes the CFT through the four-point function of the
stress tensor. Unfortunately, the four-point function is much more
difficult to analyze as it is less rigidly constrained by the
symmetries of the theory (than the two- or three-point functions)
and it depends on details of the spectrum of operators in the CFT
and their couplings to the stress tensor. As a result the four-point
function is less studied and we do not have a physical parameter to
replace the gravitational coupling $\a_3$. However, we remind the
reader that in many string constructions $\a_3\sim1/\lambda^{3/2}$
where $\lambda$ is the 't Hooft coupling in the dual superconformal
gauge theory \cite{bmps}. We also re-iterate here that at order
$C^4$ in the effective gravitational action, one could write down
five independent contractions of the Weyl tensor. Hence in complete
generality, there would be five independent gravitational couplings
appearing at this order \cite{tba2}. In our analysis, we have chosen
one particular linear combination of interactions which arises
naturally as the leading $C^4$ term in constructions of type IIB
superstring theory \cite{bmps}. While we cannot be sure that the
leading interaction at this order will have precisely the form of
$W(C)$ in \reef{act2}, we can take our results as representative of
the general case \cite{tba2}.

Hence we characterize the CFT with the physical parameters
$\{a,\delta$=$(c-a)/a,t_4,\a_3\}$. Then using \eqref{ca2} and
\eqref{t4def}, we can re-express $\{P,\frac \eta
s,\tau_\pi,{\lambda}_1\}$ as:
\begin{equation}
\begin{split}
P=&\frac{\pi^2}{2} a\, T^4\biggl\{1+\frac 94 \t+\frac 38 \t^2
+\frac{1}{180}t_4+15\a_3+\calo\left(\t^3,\t t_4,t_4^2,
\a_3^2\right)\biggr\}\,,\\
\frac{\eta}{s}=&\frac {1}{4\pi}\biggl\{1- \t +\frac 74
\t^2-\frac{4}{45} t_4+120\a_3+\calo\left(\t^3,\t t_4,t_4^2,
\a_3^2\right)\biggr\}\,,\\
\tau_\Pi T=&\frac {1}{2\pi}\biggl\{2-\ln 2-
\frac{11}{8}\t-\frac{125}{64}\t^2-\frac{13}{540}t_4
+\frac{375}{2}\a_3+\calo\left(\t^3,\t t_4,t_4^2,
\a_3^2\right)\biggr\}\,,\\
\frac{{\lambda}_1 T}{\eta}=&\frac {1}{2\pi}\biggl\{1-
\frac{1}{4}\t-\frac{73}{32}\t^2 -\frac{1}{135}
t_4+215\a_3+\calo\left(\t^3,\t t_4,t_4^2,
\a_3^2\right)\biggr\}\,.\\
\end{split}
\eqlabel{ff}
\end{equation}
We have explicitly noted that these expressions will receive higher
order corrections to remind the reader that we are still working
within a perturbative framework. In particular, consistency of the
holographic calculations requires that $a\gg1$, $\delta\ll1$ and
$t_4\ll1$, as well as $\a_3\ll1$.

In certain situations, it may be convenient to use
$\{c,\delta'\!\equiv\!(c-a)/c,t_4,\a_3\}$ to characterize the CFT
instead. In this case, \eqref{ca2} is replaced with
\begin{equation}
\frac{L^3}{\lp^3}=\frac{c}{\pi^2}\left(1-\delta'\right)\,,\qquad
\a_1=\frac{1}{8}\,\frac{\delta'}{1-\delta'}=\frac{1}{8}\,\delta'
+\frac{1}{8}\,\delta'^2+\calo(\delta'^3) \,. \eqlabel{ca3}
\end{equation}
In terms of these physical parameters $\{P,\frac \eta
s,\tau_\pi,{\lambda}_1\}$ become:
\begin{equation}
\begin{split}
P=&\frac{\pi^2}{2}c\, T^4\biggl\{1+\frac 54 \t'+\frac 38 \t'^2
+\frac{1}{180}t_4+15\a_3+\calo\left(\t'^3,\t' t_4,t_4^2,
\a_3^2\right)\biggr\}\,,\\
\frac{\eta}{s}=&\frac {1}{4\pi}\biggl\{1- \t' +\frac 34
\t'^2-\frac{4}{45} t_4+120\a_3+\calo\left(\t'^3,\t' t_4,t_4^2,
\a_3^2\right)\biggr\}\,,\\
\tau_\Pi T=&\frac {1}{2\pi}\biggl\{2-\ln 2-
\frac{11}{8}\t'-\frac{213}{64}\t'^2-\frac{13}{540}t_4
+\frac{375}{2}\a_3+\calo\left(\t'^3,\t' t_4,t_4^2,
\a_3^2\right)\biggr\}\,,\\
\frac{{\lambda}_1 T}{\eta}=&\frac {1}{2\pi}\biggl\{1-
\frac{1}{4}\t'-\frac{81}{32}\t'^2 -\frac{1}{135}
t_4+215\a_3+\calo\left(\t^3,\t' t_4,t_4^2,
\a_3^2\right)\biggr\}\,.\\
\end{split}
\eqlabel{ff2}
\end{equation}

Now with either parametrization, \reef{ff} or \reef{ff2}, these
results might be used to make a quantitative comparison with the
sQGP, as suggested in \cite{bmps,bms}. However, we should examine
different scenarios that might naturally arise in hCFT where these
results can be consistently applied for such a comparison. First,
recall that, as discussed in \cite{bms}, the gravitational couplings
are typically suppressed by the ratio of the Planck scale to the AdS
curvature scale with $\a_n\sim(\lp/L)^{2n}$. In this case, beyond
having each $\a_n\ll1$, there would be a hierarchy amongst the
couplings with $\a_{n+1}/\a_n\sim(\lp/L)^2\ll1$.

In working with the gravitational action \reef{act2}, we are
limiting our attention to the behaviour of the stress-energy in the
dual CFT and we are assuming that we can overlook the effects of any
other operators on the properties of the plasma. It was explained in
\cite{bms} that this approach is consistent up to first order in the
expansion in $(\lp/L)^2$. However, additional considerations are
required to go to higher orders when the spectrum of the CFT
includes operators with dimension of $\calo(1)$ or exactly marginal
operators, as we now describe.

Complications arise when the dual fields have linear couplings to
higher curvature terms.\footnote{Any couplings to the Einstein term
can be eliminated by a conformal transformation. For example,
$\phi^n\,R$ is removed by redefining $g_{ab}\rightarrow
\phi^{-2n/3}g_{ab}$.} For example, we might consider a coupling of
the form $\phi\,C^2$ with some massive scalar field $\phi$. In this
case, the leading order black hole solution implicitly includes
$\phi=0$. However, the scalar will acquire a nontrivial profile at
higher orders when the effects of the higher curvature terms are
included. That is, at higher orders, the dual operator acquires an
expectation value in the CFT plasma. As the hydrodynamic properties
of the plasma refer the physics at very long wavelengths, one might
attempt to proceed by integrating out this massive scalar. To be
explicit, imagine we have the scalar action
\begin{equation}
\call=-\frac{1}{2\lp^3}\left[(\nabla\phi)^2-M^2\,\phi^2+2L^2\,\b \,
\phi\, C_{abcd}C^{abcd} \right]\,, \eqlabel{r1}
\end{equation}
where $\b\sim (\lp/L)^2$, following the discussion in \cite{bms}. If
we integrate out the scalar, the contribution to the action becomes
\begin{equation}
\call=\frac{1}{2\lp^3}\left[L^6\,\frac{\b^2}{M^2L^2}\,
(C_{abcd}C^{abcd})^2 +L^8 \,\frac{\b^2}{M^4L^4}\,
(C_{abcd}C^{abcd})\, \nabla^2(C_{efgh}C^{efgh})+\cdots\right]\,.
\eqlabel{r2}
\end{equation}
Now, when the scalar has a Planck scale mass, \ie $M\sim1/\lp$, the
couplings of these higher curvature terms are suppressed in accord
with the expected hierarchy. Such a scalar would correspond to an
operator with a very large dimension, which grows parametrically
with the central charge. If instead, one considers a (scalar)
operator with dimension of $\calo(1)$, the dual field would have a
mass $M\sim 1/L$, eg, as might arise in the Kaluza-Klein reduction
of a ten-dimensional string background. In this case, coupling of
the new quartic curvature term is only suppressed by
$\beta^2\sim(\lp/L)^4$ and so this term will correct the
thermodynamic and the transport properties of the holographic plasma
at order the same order as $\a_1^2$ and $\a_2$. In fact, all of the
higher order terms in \reef{r2} have the same suppression and so can
be expected to contribute at this same order. Essentially this
demonstrates that integrating out this scalar is not an effective
approach to incorporating the effects of the dual operator with
$\calo(1)$-dimension. Hence, if we want to work with a purely
gravitational action beyond order $(\lp/L)^2$, we must impose the
absence of $\calo(1)$-dimension operators in the hCFT as a
consistency requirement. Such a condition might naturally arise in
non-supersymmetric large-$N$  gauge theories where a generic
operator develops a large anomalous dimension.

In the case of an exactly marginal (scalar) operator, the situation
is a bit more subtle. The dual scalar field $\phi^M$ is precisely
massless and so, in principle, it can have an arbitrary value in the
AdS$_5$ vacuum. Further one should think that the coupling constants
in the effective gravitational action have a(n unspecified)
dependence on this scalar, \ie $\alpha \rightarrow \alpha(\phi^M)$.
The key point then is that if $\phi^M$ becomes very large, the
couplings may not be suppressed as we had initially assumed above
\cite{bms}. This scenario naturally arises in many supersymmetric
realizations of the AdS/CFT correspondence in string theory where
the dilaton, \ie the string coupling, is dual to an exactly marginal
operator. Further, in all examples of conformal gauge theories of
which we are aware, the presence of an exactly marginal coupling
implies supersymmetry. In turn, supersymmetry would imply that
$\a_2$ vanishes, as explained in \cite{bms}, and hence the
interactions quartic in the curvatures provide the next set of
corrections in our perturbative expansion.

Hence we can identify two cases where our results can be
consistently applied in a quantitative comparison with the sQGP:

\noindent $(i)$ The CFT does not have any operators (other than the
stress energy tensor) with $\calo(1)$-dimension and in particular,
has no exactly marginal operators. In this case, one can truncate
the holographic theory to include only gravity, as in \reef{act2}.
Further, the expected hierarchy should hold amongst the
gravitational couplings, so that  $\a_2\sim \a_1^2\gg\a_3$. Hence it
is consistent to work with \reef{p1} and \reef{transport} at order
$\calo(\a_1^2,\a_2)$, while dropping the $\calo(\a_3)$
contributions.

\noindent $(ii)$ The CFT has at least one exactly marginal (scalar)
coupling, which we assume implies supersymmetry. As explained above,
$\a_2$ vanishes and we cannot necessarily assume $\a_3\ll\a_1$, in
this scenario. Hence such a case can be consistently described by
working with our results at order $\calo(\a_1,\a_3)$, while dropping
the $\a_2$ contributions.

While in either of these scenarios provides a framework in which
data might be consistently fit with our holographic results
\reef{ff}, neither one seems a particulary good conjecture as to the
type of CFT which might describe the sQGP. However, one can
certainly imagine other interesting situations. For example, there
may be CFT's where $\beta$ vanishes or has some accidental
suppression. This would reflect an unexpected suppression of the
correlator of the dual operator and two stress
tensors.\footnote{This suppression might be achieved by some
underlying symmetry, \eg $\phi\rightarrow-\phi$.} Hence one might
also explore less cautious comparisons without restricting to the
two scenarios above \cite{tba2}.

What are the prospects of further developing the holographic model
of sQGP? In the framework of holographic conformal models, a natural
venue to pursue is the to go even higher order in all the couplings,
ultimately  to finite values of $\{a,\dd,t_4,\cdots\}$, rather than
restricting\footnote{Note that these physical parameters cannot take
completely arbitrary values, as they are constrained by the
consistency of the underlying CFT \cite{hm,relax,diego,vi6}.} to the
`Einstein gravity corner' where the higher curvature couplings are
all small, \ie $a\gg1$, $\delta\ll1$ and $t_4\ll1$, as well as
$\a_3\ll1$. The advantage of this approach is that conformal
invariance severely constrains thermal and transport properties of
the plasma, compared to the proliferation of the transport
coefficients in non-conformal theories \cite{rom}. As we discussed
above, the challenge here is that one must expand the dual
gravitational theory to include additional fields which can account
for the effects of the condensates of various $\calo(1)$-dimension
operators. Alternatively, one can stay in the linearized
approximation of the large-$N$, strong coupling supersymmetric CFT
plasma (as in \cite{bms}), but include in addition (small)
corrections due to breaking of the scale invariance. To leading
order in the hydrodynamics approximation, this introduces a new
viscous coefficient: the bulk viscosity. While the bulk viscosity is
rather well studied in holographic models \cite{bulk}, much less is
known about the second order transport coefficients of non-conformal
theories -- see \cite{ns} for some initial investigations. Either
way, the utility of further holographic models of sQGP, as well as
hydrodynamic plasma simulations \cite{sym}, rests upon ability to
extract additional observables from RHIC and future LHC experiments.

\section*{Acknowledgments}
It is a pleasure to thank Aninda Sinha for useful conversations.
Research at Perimeter Institute is supported by the Government of
Canada through Industry Canada and by the Province of Ontario
through the Ministry of Research \& Innovation. AB gratefully
acknowledges further support by an NSERC Discovery grant and support
through the Early Researcher Award program by the Province of
Ontario. RCM also acknowledges support from an NSERC Discovery grant
and funding from the Canadian Institute for Advanced Research.
During the course of work, MH has been supported by Polish Ministry
of Science and Higher Education under contracts N N202 247135 and N
N202 105136 and Foundation for Polish Science START Programme. MH
also acknowledges hospitality of Perimeter Institute for Theoretical
Physics and Weizmann Institute of Science.

\end{document}